\documentclass[%
 preprint,
 superscriptaddress,
 amsmath,amssymb,
 aps,
 showkeys,
 titlepage,
 endfloats*.   
]{revtex4-2}

\usepackage[utf8]{inputenc}
\usepackage[english]{babel}

\usepackage{graphicx}
\usepackage{dcolumn}
\usepackage{bm}
\usepackage[mathlines]{lineno}

\usepackage[colorlinks = true,
            linkcolor = blue,
            urlcolor  = blue,
            citecolor = red,
            anchorcolor = blue]{hyperref}
\usepackage{color, soul}
\usepackage{comment}
\usepackage{natbib}
\newcommand*{\rom}[1]{\expandafter\@slowromancap\romannumeral #1@}

\usepackage[version=4]{mhchem}
\newcommand{\code}[1]{\texttt{#1}}
\usepackage{chemformula}
\usepackage{threeparttable}
\usepackage{hhline}
\usepackage{dsfont}

\begin{document}


\title{Absence of Phonon Softening across a Charge Density Wave Transition due to Quantum Fluctuations } 

\author{Yubi Chen}
\affiliation{%
 Department of Physics, University of California, Santa Barbara, California 93106-9530, USA}%
\affiliation{Department of Mechanical Engineering, University of California, Santa Barbara, CA 93106-5070, USA}

\author{Terawit Kongruengkit}
\affiliation{Materials Department, University of California, Santa Barbara, California 93106-5050, USA}

\author{Andrea Capa Salinas}
\affiliation{Materials Department, University of California, Santa Barbara, California 93106-5050, USA}

\author{Runqing Yang}
\affiliation{Department of Mechanical Engineering, University of California, Santa Barbara, CA 93106-5070, USA}

\author{Yujie Quan}
\affiliation{Department of Mechanical Engineering, University of California, Santa Barbara, CA 93106-5070, USA}

\author{Fanghao Zhang}
\affiliation{Department of Mechanical Engineering, University of California, Santa Barbara, CA 93106-5070, USA}

\author{Ganesh Pokharel}
\affiliation{Materials Department, University of California, Santa Barbara, California 93106-5050, USA}
\affiliation{Perry College of Mathematics, Computing, and Sciences, University of West Georgia, Carrollton, GA 30118}

\author{Linus Kautzsch}
\affiliation{Materials Department, University of California, Santa Barbara, California 93106-5050, USA}

\author{Sai Mu}
\affiliation{%
 SmartState Center for Experimental Nanoscale Physics, Department of Physics and Astronomy, University of South Carolina, SC 29208, USA}%

\author{Stephen D. Wilson}
\affiliation{Materials Department, University of California, Santa Barbara, California 93106-5050, USA}


\author{John W. Harter}
\email{harter@ucsb.edu} 
\affiliation{Materials Department, University of California, Santa Barbara, California 93106-5050, USA}

\author{Bolin Liao}
\email{bliao@ucsb.edu} 
\affiliation{Department of Mechanical Engineering, University of California, Santa Barbara, CA 93106-5070, USA}


\begin{abstract}

Kagome metals have emerged as a frontier in condensed matter physics due to their potential to host exotic quantum states. 
Among these, \ce{CsV3Sb5} has attracted significant attention for the unusual coexistence of charge density wave (CDW) order and superconductivity, presenting an ideal system for exploring novel electronic and phononic phenomena. 
The nature of CDW formation in \ce{CsV3Sb5} has sparked considerable debate.
Previous studies have suggested that the underlying mechanism driving the CDW transition in \ce{CsV3Sb5} is distinct from conventional ones, such as electron-phonon coupling and Fermi surface nesting.
In this study, we examine the origin of the CDW state via \textit{ab initio} finite-temperature simulations of the lattice dynamics associated with CDW structures in \ce{CsV3Sb5}.
Through a comparative study of \ce{CsV3Sb5} and 2H-\ce{NbSe2}, we demonstrate that the experimental absence of phonon softening in \ce{CsV3Sb5} and the presence of a weakly first order transition can be attributed to quantum zero-point motion of the lattice, which leads to smearing of the CDW landscape and effectively stabilizes the pristine structure even below the CDW transition temperature. 
We argue that this surprising behavior could cause coexistence of pristine and CDW structures across the transition and lead to a weak first-order transition. 
We further discuss experimental implications and use the simulation to interpret coherent phonon spectroscopy results in single crystalline \ce{CsV3Sb5}. 
These findings not only refine our fundamental understanding of CDW transitions, but also highlight the surprising role of quantum effects in influencing macroscopic properties of relatively heavy-element materials like \ce{CsV3Sb5}.
Our results provide crucial insights into the formation mechanism of CDW materials that exhibit little to no phonon softening, including cuprates, aiding in the understanding of the CDW phase in quantum materials.

\end{abstract}

                            
\maketitle


\section{Introduction}

Characterized by corner-sharing triangles~\cite{syozi1951statistics}, Kagome metals have become a focal point in condensed matter physics due to inherent features such as geometric frustration, flat bands, Dirac fermions, and van Hove singularities (VHS) with different electron fillings~\cite{kiesel2013unconventional,wang2013competing}.
These features enable the potential to host exotic quantum phenomena~\cite{ko2009doped,guo2009topological,o2010strongly,mazin2014theoretical} such as
unconventional superconductivity~\cite{yu2012chiral,wu2021nature,lin2022multidome,oey2022fermi}, 
anomalous Hall effect~\cite{yang2020giant,yu2021concurrence}, 
spin liquid states~\cite{yan2011spin,han2012fractionalized}
and quantum magnetism~\cite{han2012fractionalized,yin2020quantum}, etc.
Among Kagome materials, the newly discovered $\mathds{Z}_2$ topological metal \ce{CsV3Sb5}~\cite{ortiz2019new,ortiz2020cs,ortiz2021fermi,wilson2024av3sb5} stands out due to its quasi two-dimensional structure, correlated electron effects, and absent local spin moments~\cite{ortiz2019new,ortiz2020cs}.
\ce{CsV3Sb5} is especially intriguing for the coexistence of 
 a charge density wave (CDW) order below 94~K and superconductivity below 2.5~K~\cite{jiang2021unconventional,liang2021three,yu2021unusual,chen2021double}.
Understanding the nature of its CDW properties is not only crucial for fundamental physics~\cite{mielke2022time,park2021electronic} 
but also important for developing next-generation quantum devices~\cite{wang2023anisotropic}.
Nevertheless, the precise mechanism driving the CDW order in \ce{CsV3Sb5}remains poorly understood, presenting a substantial challenge both theoretically and experimentally.

Previous research has suggested that conventional factors driving CDW transitions, such as Fermi surface nesting and strong electron-phonon coupling (EPC) may not play a major role in \ce{CsV3Sb5}~\cite{johannes2006fermi,johannes2008fermi,zhu2015classification,zhu2017misconceptions}. 
For example, even though angle-resolved photoemission spectroscopy (ARPES) experiments~\cite{kang2022twofold,hu2022rich} observed a few VHSs close to the Fermi level, which forms Fermi surface nesting to potentially drive a CDW transition, calculations of the Lindhard susceptibility~\cite{kaboudvand2022fermi,wang2022origin,si2022charge} have shown that Fermi surface nesting is not the predominant cause, as the nesting wavevector does not coincide with the CDW wavevector. 
Furthermore, \ce{CsV3Sb5} differs strongly from EPC-driven CDW materials, like \ce{NbSe2}~\cite{zhu2015classification}, because Raman and X-ray measurements of \ce{CsV3Sb5} phonons do not exhibit the expected phonon softening behavior across the CDW transition~\cite{li2021observation,liu2022observation} — a hallmark of EPC-induced CDW materials~\cite{zhu2015classification,zhu2017misconceptions}, despite measurements and simulations showing moderate EPC strengths~\cite{zhong2023testing,xie2022electron,si2022charge,tan2021charge,zhang2021first}.
Explanations have been suggested for the absence of phonon softening, such as a very broad linewidth of the soft phonon mode~\cite{gutierrez2023phonon} and a persistent phason gap~\cite{miao2021geometry}, but the confirmation of these conjectures remains elusive.
Alternative mechanisms for CDW formation have been proposed, such as a Jahn-Teller-like transition~\cite{wang2022origin}, exciton condensation (electron-hole coupling)~\cite{van2010exciton,chen2018reproduction} etc.
These approaches have not conclusively explained all the experimental observations, suggesting that additional factors might be at play.

In this study, we use \textit{ab initio} simulations to carefully analyze the energy landscape in \ce{CsV3Sb5} associated with various CDW distortions. We then employ a finite-temperature lattice dynamics technique based on the stochastic self-consistent harmonic approximation (SSCHA)~\cite{monacelli2021stochastic} to examine how thermal and quantum zero-point fluctuations effectively modify the energy landscape and impact the phonon dynamics in \ce{CsV3Sb5}.
This approach (see Supplemental Material~\cite{SM} for more details) intends to go beyond a previous study using a temperature-invariant perturbative potential without considering the quantum zero-point effect~\cite{ptok2022dynamical}.
%
Through a detailed comparative study with 2H-\ce{NbSe2}, we elucidate the important role of quantum zero-point motion in suppressing the phonon softening behavior in \ce{CsV3Sb5}.
In addition to stabilizing the \ce{CsV3Sb5} pristine phase, quantum fluctuations also reduce the energy barriers that separate metastable CDW states, leading to a weak first-order CDW transition and the simultaneous presence of CDW and pristine phases around the transition temperature.
Accordingly, we propose that a dynamic view of the free-energy landscape, allowing for the coexistence of metastable CDW states, could emerge at lower temperatures, leading to the phonon linewidth broadening observed in our coherent phonon spectroscopy (CPS) of single-crystalline \ce{CsV3Sb5}.
Our result challenges the prevailing understanding of phonon softening behavior as a necessary indicator of EPC-induced CDW transitions.
This work reconciles the possibility of EPC as the origin of CDW  with the experimental absence of phonon softening in \ce{CsV3Sb5}, with implications that can be generalized to other unconventional CDW materials like cuprates~\cite{zhu2015classification,zhu2017misconceptions}.
Hence, it paves the way for reinterpreting previous experimental observations and for future quantum applications of these materials. 
Moreover, we want to highlight the surprising role of quantum mechanical zero-point effects in a heavy-element material like \ce{CsV3Sb5}. This resembles the quantum paraelectricity in the heavy-element \ce{SrTiO3}, where the ferroelectric order is also suppressed by quantum fluctuations~\cite{muller1979srti,verdi2023quantum}.


\section{Methods}

\subsection{Computational Methods}

First-principles density functional theory (DFT) calculations were conducted using the Vienna Ab initio Simulation Package (\code{VASP}) version 6.4.1~\cite{kresse1996vasp1, kresse1996vasp2}, using the projector augmented wave (PAW) pseudopotentials~\cite{blochl1994projector}. 
A 300~eV plane-wave cutoff was employed for \ce{CsV3Sb5}, and 400~eV cutoff for \ce{NbSe2}. 
The exchange-correlation functional was chosen to be the Perdew–Burke–Ernzerhof (PBE) functional~\cite{perdew1996generalized} with van der Waals (vdW) D3 correction~\cite{grimme2011effect}, denoted as PBE+vdW. 
The electronic configurations for the elemental pseudopotentials were: Cs ($5s^26p^66s^1$), V ($3s^23p^63d^34s^2$), Sb ($5s^25p^3$), Nb ($4p^64d^35s^2$), and Se ($4s^24p^4$). 
Brillouin zone sampling was performed using an 18$\times$18$\times$12 $\Gamma$-centered \textbf{k} grid for \ce{CsV3Sb5} unit cell, and a 24$\times$24$\times$8 $\Gamma$-centered \textbf{k} grid for 2H-\ce{NbSe2} unit cell. 
The converged lattice parameters were $a=b=5.4491$~\AA, $c=9.3056$~\AA~ for pristine \ce{CsV3Sb5} and $a=b=3.4517$~\AA, $c=12.3909$~\AA~ for pristine 2H-\ce{NbSe2}, in a good agreement to experimental data 
$a=b=5.52$~\AA, $c=9.36$~\AA~for \ce{CsV3Sb5}~\cite{ortiz2020cs} and 
$a=b=3.4446$~\AA, $c=12.5444$~\AA~for 2H-\ce{NbSe2} at 298~K~\cite{marezio1972crystal}.
Both systems utilized no spin polarization for the absent local spin moments~\cite{ortiz2019new,ortiz2020cs} and a Gaussian smearing of 0.1~eV.
The convergence threshold was set as $1\times10^{-7}$~eV for energy and as 0.005~eV/\AA~for force.
The harmonic phonon calculations were executed using the finite-displacement method in Phonopy~\cite{togo2015first}, with supercell sizes of 3$\times$3$\times$2 for \ce{CsV3Sb5} (162 atoms) and 6$\times$6$\times$1 for 2H-\ce{NbSe2} (216 atoms), using $\Gamma$-centered supercell \textbf{k} grids of 3$\times$3$\times$3 and 2$\times$2$\times$2, respectively. We found that the harmonic phonon dispersion in \ce{CsV3Sb5} was strongly affected by the supercell size, whose convergence needs to be carefully checked (See Fig.~S1 of of Supplemental Material~\cite{SM} for the supercell convergence in \ce{CsV3Sb5} pristine phonon).
Spin-orbit coupling (SOC) was turned off after verifying its negligible effect on phonon dispersions (See Fig.~S2 of Supplemental Material~\cite{SM} for a comparison of pristine phonon dispersions with PBE+vdW+SOC and PBE+vdW).

The SSCHA techinique was employed to explore anharmonic and quantum effects on lattice dynamics at a nonperturbative level~\cite{errea2014anharmonic,bianco2017second,monacelli2018pressure} (See the Supplemental Material~\cite{SM} for more details).
The SSCHA method variationally minimizes the free energy by optimizing lattice parameters and internal coordinates while preserving the symmetry of the system. Gaussian ensembles of lattice distortions surrounding an average structure are sampled to estimate the lattice free energy, which is then minimized by adjusting the average structure and the spread of the Gaussian probability distribution including both thermal and quantum effects.
Self-consistency was achieved for each temperature by utilizing 200 configurations per ensemble, ensuring that the error ratio to the free-energy gradient stayed below $1\times 10^{-4}$.
Once self-consistency was achieved, phonon dispersions at finite temperatures were calculated using the free energy Hessians~\cite{bianco2017second} evaluated with up to 5000 configurations.
The computational setup of ensemble configurations was aligned with the harmonic phonon calculations, maintaining consistency across our methodology.


\subsection{Experimental Methods}

For the CPS experiment, optical pump-probe transient reflectivity measurements were performed on a freshly-cleaved sample of \ce{CsV3Sb5} mounted in an optical cryostat. 
A noncollinear optical parametric amplifier generated $\sim$70~fs signal (800~nm) and idler (1515~nm) pulses at a 50~kHz repetition rate, with a pump fluence of approximately 100~$\mathrm{mJ/cm^2}$.
Measurements at different temperatures were performed upon successive warming from a base temperature of 9~K. 
To isolate the coherent phonon oscillations, an exponential background was subtracted from the transient change in reflectivity, followed by an apodization step, and Fourier transforming to the frequency domain. 
Further details on the experimental technique can be found in Ref.~\cite{ratcliff2021coherent}.

\section{Results and Discussions}

\subsection{Conventional Phonon Behavior of EPC-driven CDW Materials}
\begin{figure}[!thb]
\includegraphics[width=1.0\textwidth]{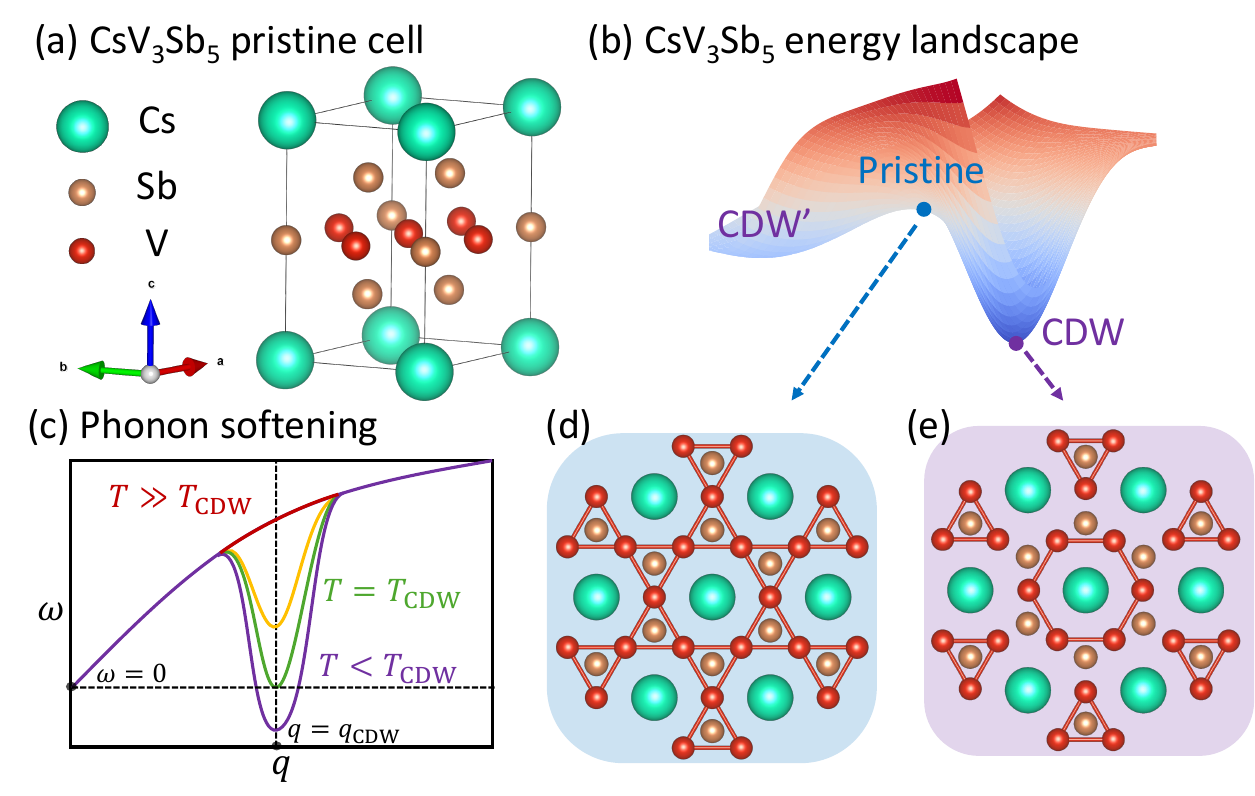}
\caption{
(a) The atomic structure of \ce{CsV3Sb5} pristine cell. 
(b) A schematic of \ce{CsV3Sb5} energy landscape,  featuring multiple potential wells representing the CDW structures and a saddle point representing the pristine structure. 
(c) A schematic illustration of typical phonon softening behavior for the pristine structure across the CDW transition temperature $T_\mathrm{CDW}$. 
The phonon mode corresponding to the CDW structure ($q=q_\mathrm{CDW}$) is gradually destabilized with decreasing temperatures.
Notably, \ce{CsV3Sb5} does not exhibit the phonon softening behavior.
(d) Top-view of \ce{CsV3Sb5} pristine structure with symmetry $P6/mmm$, where V atoms form Kagome lattice. 
The V-V pairs are connected with the same bond length 2.725~\AA.
(e) Top-view of the  \ce{CsV3Sb5} tri-hexagonal (TrH) layers.
The V-V bond distance cutoff is set as 2.7~\AA, and pairs of V atoms that exceed this distance are not connected.
} 
\label{fig:1-1}
\end{figure}

Figure~\ref{fig:1-1}(a) illustrates the pristine cell of \ce{CsV3Sb5}, featuring a layered structure composed of alternating atomic planes. 
The vanadium (V) atoms are arranged in a two-dimensional Kagome lattice, surrounded by hexagonal antimony (Sb) atoms. 
The Kagome plane is sandwiched between two honeycomb Sb layers, with cesium (Cs) atoms situated between the V-Sb layers. 
CDW materials typically transition from a high-symmetry pristine structure to a lower-symmetry CDW structure as the temperature decreases. 
For instance, \ce{CsV3Sb5} transitions from a $P6/mmm$ structure at room temperature, depicted in Fig.~\ref{fig:1-1}(d), to a CDW structure at 94~K, which is the tri-hexagonal (TrH) 2$\times$2$\times$2  or mixed TrH star-of-David (SoD) 2$\times$2$\times$4 configuration.~\cite{ortiz2021fermi,li2021observation,stahl2022temperature} with an in-plane lattice shown in Fig.~\ref{fig:1-1}(e).
Similarly, 2H-\ce{NbSe2} undergoes a transition into its CDW structure at 33~K~\cite{weber2011extended}. (See Supplemental Material~\cite{SM} for more details about the CDW order in 2H-\ce{NbSe2}.~\cite{zheng2018first,silva2016electronic,guster2019coexistence})
From static ground-state DFT, conducted at zero temperature, the CDW structures are energetically favorable compared to the pristine structure.
Considering a continuous variation of structural coordinates, where each structure correlates with one energy, we can visualize an energy landscape, as illustrated in Fig.~\ref{fig:1-1}(b). 
In this energy landscape, the potential wells correspond to the energetically stable CDW structures, and the saddle point indicates the unstable pristine structure. 

EPC-driven CDW transitions are often accompanied by a phonon softening behavior sketched in Fig.~\ref{fig:1-1}(c)\cite{weber2011extended,weber2011electron}.
Due to the unstable saddle point in the energy landscape, the pristine structure at zero temperature should display imaginary phonon frequencies at phonon wavevectors corresponding to the CDW distortions, as depicted by the purple curve in Fig.~\ref{fig:1-1}(c). 
As the temperature increases, the imaginary phonon is gradually suppressed, which becomes stable ($\omega \geq 0$) above the CDW transition temperature ($ T \geq T_\mathrm{CDW}$). 
For instance, experimental observations of EPC-driven CDW in 2H-\ce{NbSe2} show these typical softening behaviors illustrated in Fig.~\ref{fig:1-1}(c).~\cite{weber2011extended}
In contrast, \ce{CsV3Sb5} does not follow this pattern: Experimental results reveal no phonon softening~\cite{li2021observation,liu2022observation}, which was previously interpreted to suggest that EPC is not the dominant mechanism driving the CDW transition in \ce{CsV3Sb5}.

\begin{figure}[!thb]
\includegraphics[width=1.0\textwidth]{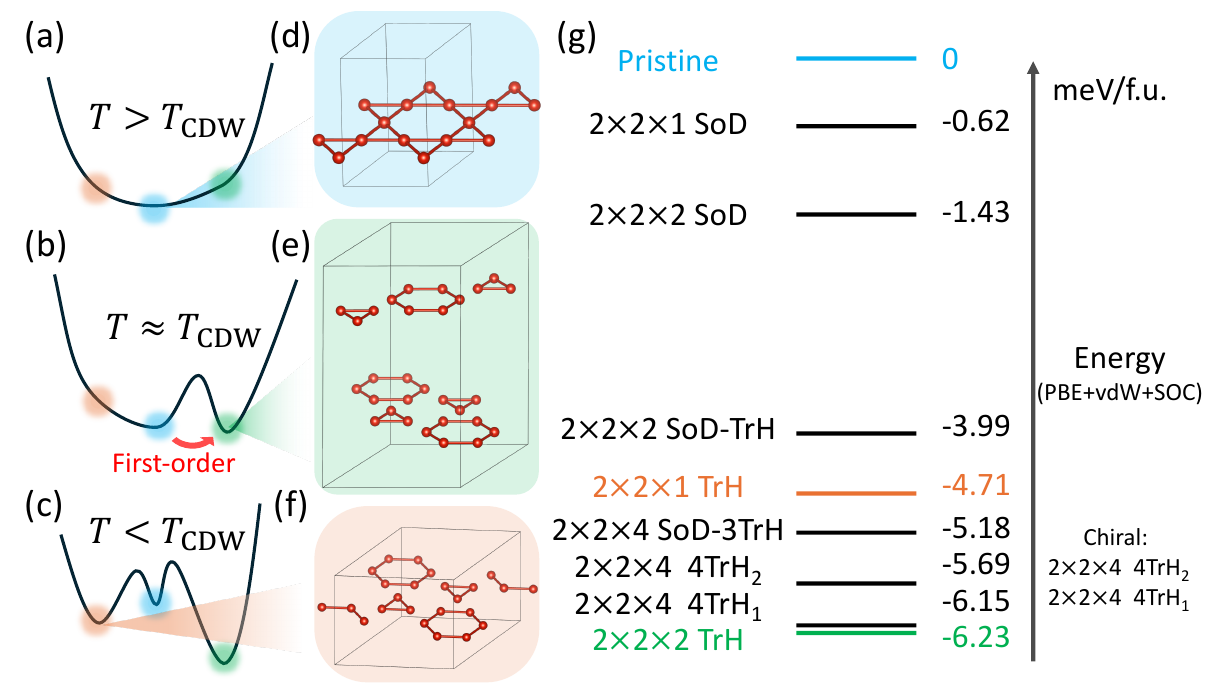}
\caption{
The schematic evolution of the free energy landscape at (a) above CDW transition temperature ($T>T_\mathrm{CDW}$) (b) near transition temperature ($T \approx T_\mathrm{CDW}$) and (c) below transition temperature ($T < T_\mathrm{CDW}$).
(d) The pristine structure of \ce{CsV3Sb5}.
(e) The CDW order with 2$\times$2$\times$2 TrH layers
(f) The CDW order with 2$\times$2$\times$1 TrH layers
(g) The energy hierarchy of identified \ce{CsV3Sb5} CDW orders in the units of meV per formula unit (meV/f.u.).
} 
\label{fig:1-2}
\end{figure}

Figure~\ref{fig:1-2} (a,b,c) illustrates the schematic evolution of the free energy landscape in \ce{CsV3Sb5} as the temperature decreases. 
In Fig.~\ref{fig:1-2}(a), above the CDW transition temperature ($T>T_\mathrm{CDW}$), the pristine structure, which is labeled by the blue dot with the atomic structure shown in Fig.~\ref{fig:1-2}(d), represents a global energy minimum and the system favors the high-symmetry pristine state.
As the temperature approaches the CDW transition ($T \approx T_\mathrm{CDW}$) in Fig.~\ref{fig:1-2}(b), low-symmetry CDW structures, like the 2$\times$2$\times$2~TrH structure shown in Fig.~\ref{fig:1-2}(e) indicated by the green dot, exhibit free energies comparable to that of the pristine structure, facilitating the onset of the CDW transition.
Below the transition temperature ($T < T_\mathrm{CDW}$) in Fig.~\ref{fig:1-2}(c), additional CDW orders, such as the orange dot 2$\times$2$\times$1~TrH order in Fig.~\ref{fig:1-2}(f), become local minima with lower free energies than the pristine phase.
The rationale for depicting the pristine structure as a local minimum at all temperatures will be discussed in the subsequent section.

We systematically explored and identified multiple possible CDW structures in \ce{CsV3Sb5}.
Figure~\ref{fig:1-2}(g) provides an overview of the energy hierarchy of identified CDW structures as compared to the pristine structure, computed using the PBE+vdW+SOC method.
The energies of the structures are directly comparable, as they were calculated using consistent setups, with computational details and their atomic structures provided in the Supplemental Material~\cite{SM}.
Among these, the 2$\times$2$\times$2~TrH order is the most energetically favorable structure with a stable phonon dispersion.
The 2$\times$2$\times$4 4TrH$_1$ and 4TrH$_2$ chiral structures,  both consisting of four TrH layers, exhibit energies close to the global minimum, which can potentially explain the chirality observed in experiments~\cite{elmers2024chirality}.
The 2$\times$2$\times$4 SoD-3TrH structure, comprising one layer of SoD and three layers of TrH, corresponds to the structure reported in  Ref.~\cite{kautzsch2023structural}.
The 2$\times$2$\times$1 TrH structure exhibit a stable phonon dispersion, similar to the global minimum (See Supplemental Material~\cite{SM} for the stable phonon dispersions of 2$\times$2$\times$1 and 2$\times$2$\times$2 TrH orders~\cite{liu2022observation,wang2021charge,subedi2022hexagonal}).
The structures with SoD layers are found to be more unstable. 
The 2$\times$2$\times$1 SoD is dynamically unstable with imaginary phonon frequencies.
The 2$\times$2$\times$2 SoD-TrH, composed of one SoD layer and one TrH layer, and 2$\times$2$\times$2 SoD, composed of two SoD layers, only emerge under less stringent convergence criteria, as they tend to relax back to the 2$\times$2$\times$2 TrH configuration under tighter conditions.

Overall, Fig.~\ref{fig:1-2}(g) also provides an estimation of the energy barriers between the distinct structures.
The energy barrier between ground state structure and the pristine phase is about 6.23~meV per formula unit (f.u.), which corresponds to 49.84~meV($=6.23\times 8$) for the degree of freedom associated with the distortion into a 2$\times$2$\times$2 CDW structure. 
Given that the CDW transition occurs at 94~K~($\approx 8$~meV), this barrier is unlikely to be overcome solely by thermal fluctuations.
In the following section, we will show that the energy barrier could be reduced due to the effects of zero-point motion.

\subsection{Quantum Zero Point Motion}

To effectively grasp the impact of finite temperatures on CDW materials, it is crucial to incorporate both thermal fluctuations and quantum zero-point effects on their lattice dynamics, for which we employ the SSCHA method.~\cite{monacelli2021stochastic} 
Specifically, the SSCHA method aims to variationally minimize the free energy $F=E-TS$, where $E$ represents the energy and $TS$ denotes the entropy contribution~\cite{errea2013first}.
During the minimization process, SSCHA generates a set of ensemble configurations based on statistical fluctuations involving both the thermal and zero-point effects.

Zero-point motion in a lattice arises from the quantum mechanical effect that introduces position uncertainty.
The Hamiltonian of the lattice can be transformed to a set of independent harmonic oscillators. 
For a single harmonic oscillator at zero temperature, the position uncertainty of the ground state is given by $\langle x_i^2 \rangle = \frac{\hbar}{2 m \omega_i } $, where $m$ is the effective mass of a particle in a harmonic potential with frequency $\omega_i$ and $x_i$ is the displacement amplitude relative to the coordinates of the reference structure, which is the pristine structure in our case.
The mass-weighted distance from the pristine structure is $ \sqrt{ \langle m x_i^2 \rangle } =  \sqrt{ \frac{\hbar}{2\omega_i}  }$, which depends solely on the vibrational frequencies.
To simulate the quantum effects and thermal effects, we can generate an ensemble of configurations, where position $x_i$ is drawn from a Gaussian distribution, determined by the position uncertainty at finite temperatures.
By evaluating the energy of the ensemble configurations via DFT, we can extract various ensemble-averaged physical properties, such as the finite-temperature average structure, phonon dispersions, etc.


\begin{figure}[!thb]
\includegraphics[width=0.85\textwidth]{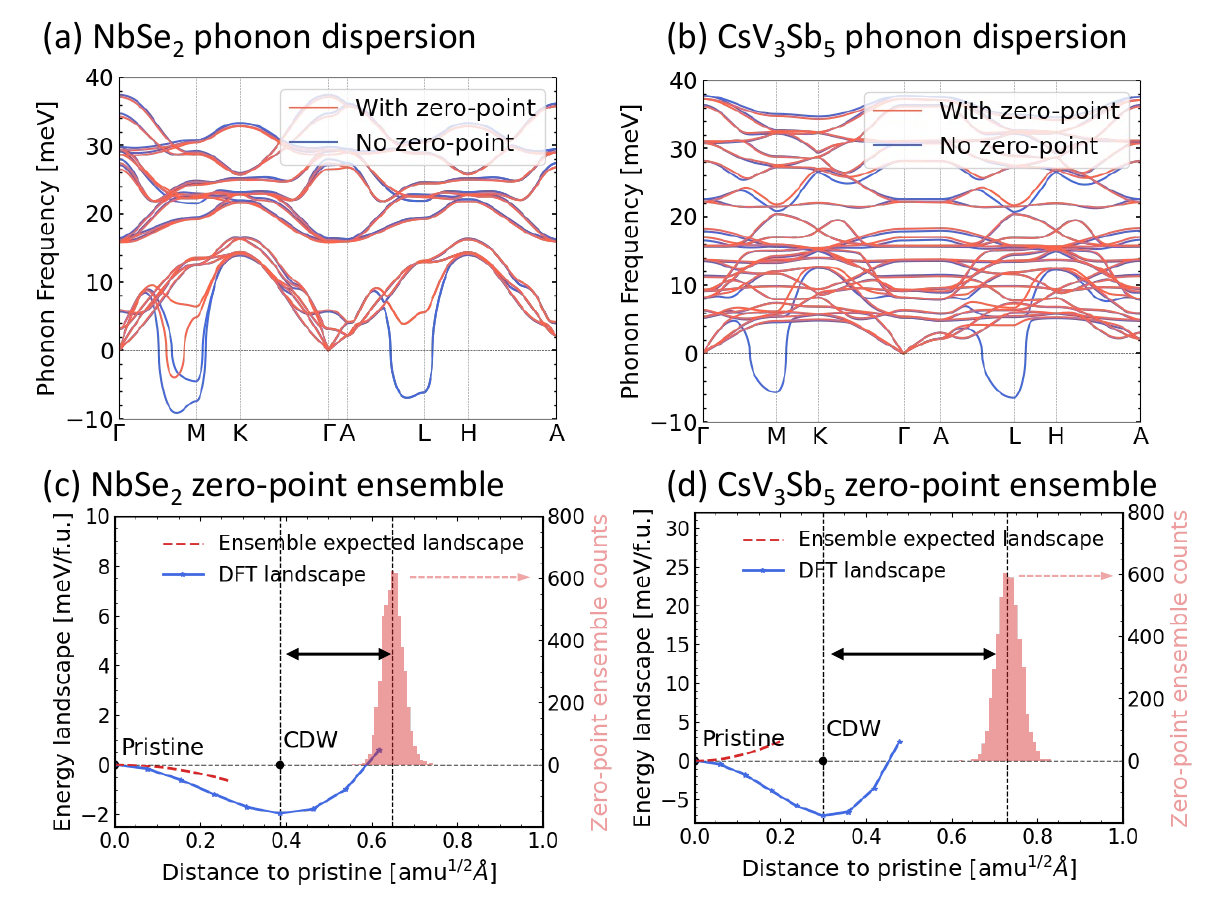}
\caption{
(a) The phonon dispersion for the pristine 2H-\ce{NbSe2} structure calculated without zero-point motion (``No zero-point") by finite-displacement methods and with zero-point motion at 5~K (``With zero-point") by SSCHA. 
(b) Similar to panel (a) but for \ce{CsV3Sb5}.
The stabilized phonon at 5~K suggests that the \ce{CsV3Sb5} pristine structure can not exhibit phonon softening behavior due to zero-point motion. 
(c) the blue dotted line depicts the DFT energy landscape of 2H-\ce{NbSe2} (left y-axis, energy per formula unit). The horizontal axis ``distance to pristine," defined as $\sqrt{\sum_i M_i (R_i - r_i)^2 / N}$, indicates the deviation of structures from the pristine cell. The right y-axis illustrates the distribution of 5000 structures in the zero-point ensemble. 
The expected free energy landscape for sampling this ensemble is sketched by a red dashed line, showing a smeared negative parabola. 
(d) Similar to panel (c) but for \ce{CsV3Sb5}. 
The red dashed line reveals a positive parabola because the \ce{CsV3Sb5} CDW potential well is not sufficiently broad to be captured by the ensemble.
} 
\label{fig:1-3}
\end{figure}

The analysis of phonon softening in 1H-\ce{NbSe2} and 2H-\ce{NbSe2} including both thermal and zero-point fluctuations by SSCHA has been performed before~\cite{bianco2020weak}. 
Here, we first replicated the phonon dispersion calculations for 2H-\ce{NbSe2}, as shown in Fig.~\ref{fig:1-3}(a). 
The blue curve represents the phonon dispersion without considering zero-point motion (``No zero-point'') calculated using the finite displacement method and its displacement amplitude was set as 0.01~\AA. 
In contrast, the red curve illustrates the phonon dispersion with zero-point motion at 5~K (``With zero-point''), performed via SSCHA free-energy Hessian calculations. 
For the 5~K calculation with zero-point motion, thermal effects are negligible, yielding results equivalent to those at 0~K, and we selected 5~K to avoid potential numerical divergence issues.
Our results were consistent with the previous study using SSCHA~\cite{bianco2020weak}, where the imaginary frequency of ``With zero-point'' is around -5~meV near $M$ point, validating our approach and computational setup.

Next, we applied the same procedure to 
\ce{CsV3Sb5}. 
Surprisingly, the 5~K ``With zero-point'' phonon dispersion exhibited dynamic stability without imaginary phonon frequencies, as shown in Fig.~\ref{fig:1-3}(b), which is in sharp contrast to the case in 2H-\ce{NbSe2}.
Figure~\ref{fig:1-3}(b) differs from another SSCHA study~\cite{gutierrez2023phonon}, which we believe is most likely due to the use of a different supercell size.
To understand the difference between 2H-\ce{NbSe2} and \ce{CsV3Sb5}, we examined the optimized ensembles used for computing the free-energy Hessian matrices and observed significant differences in the distributions of the 5000 ensemble configurations used in both calculations. 
As depicted in Fig.~\ref{fig:1-3}(c) and Fig.~\ref{fig:1-3}(d), the blue dotted line represents the energy landscape along the CDW distortion from the pristine structure calculated using DFT, with energy per formula unit referenced to the pristine structure. 
The CDW landscape is chosen to represent the most stable CDW structure for \ce{CsV3Sb5} and 2H-\ce{NbSe2}, respectively (See Supplemental Material~\cite{SM} for more details).
The horizontal axis, ``distance to pristine," is defined as the averaged mass-weighted distance to the pristine structure $\sqrt{\sum_i M_i (R_i - r_i)^2 / N}$, where $N$ is the number of atoms in a given structure, $M_i$ is the mass of the $i$-th atom in atomic mass units (amu), $R_i$ is the Cartesian coordinate of the $i$-th atom in the given distorted structure, $r_i$ is the Cartesian coordinate of the corresponding atom in the pristine cell. 
This value quantifies the normalized radial deviation of a given structure from its pristine state. 
The right y-axis illustrates the distribution of zero-point ensemble configurations based on their radial distances to the pristine structure.

The ``No zero-point'' phonon dispersion, calculated by 0.01~\AA~ finite-displacement method, probes the immediate vicinity of the pristine cell within the DFT landscape. 
In contrast, the ensemble due to zero-point motion detects a broader range of distances extending beyond the CDW potential well in Fig.~\ref{fig:1-3}(c) and Fig.~\ref{fig:1-3}(d).
The first vertical dashed line of each figure denotes the location of the CDW potential well, whereas the second vertical dashed line indicates the average distance of configurations from the pristine structure in the optimized zero-point ensemble. 
The black double arrow represents the distance between these two vertical dashed lines and is almost twice larger in \ce{CsV3Sb5} compared to that in 2H-\ce{NbSe2}.
The red dashed lines denote the effective free-energy landscapes as a result of sampling the zero-point ensemble.
The red dashed line in Fig.~\ref{fig:1-3}(c) reveals an effective landscape with a smeared negative parabola for pristine 2H-\ce{NbSe2}, suggesting that the zero-point ensemble effectively samples the CDW potential well. In other words, even though the effective free energy landscape in 2H-\ce{NbSe2} is smeared by the position uncertainty due to zero-point motions, the pristine structure remains dynamically unstable with imaginary phonon modes. 
In contrast, Fig.~\ref{fig:1-3}(d) reveals an effective landscape with a positive parabola for \ce{CsV3Sb5},
since the average distance from the pristine structure of the zero-point ensemble is much wider than the CDW potential well, In this case, the CDW potential well cannot be effectively sampled by the ensemble, resulting in stabilized phonons with the pristine structure even near zero temperature. 
As a result, the uncertainty of the atomic position introduced by zero-point motion effectively stabilizes the \ce{CsV3Sb5} phonons in the pristine structure, leading to the absence of phonon softening at finite temperatures.

To further investigate why the absence of phonon softening is observed specifically in \ce{CsV3Sb5}, we examine the relationship between the zero-point ensemble and the CDW potential well. 
The spread of the zero-point distribution can be estimated using the quantum uncertainty relation, approximately $\sqrt{\frac{\hbar}{2\omega}}$, where $\omega$ represents the effective frequency of phonon modes $\omega_i$
(see Supplemental Material~\cite{SM} for a simplified model of the zero-point distribution).
By examining the phonon dispersions, we can assess the quantum fluctuations in the pristine phases. 
Given that the phonon dispersions in 2H-\ce{NbSe2} and \ce{CsV3Sb5} are similar, the zero-point ensemble spreads are comparable: 0.65~$\mathrm{amu}^{1/2}$\AA~for 2H-\ce{NbSe2} and 0.73~$\mathrm{amu}^{1/2}$\AA~for \ce{CsV3Sb5}. 
What distinguishes these two cases is the relative distances of the CDW structures to the pristine structures. 
In \ce{NbSe2}, the CDW structure is located at a distance 0.39~$\mathrm{amu}^{1/2}$\AA~ from pristine, while in \ce{CsV3Sb5}, this distance is shorter, at 0.30~$\mathrm{amu}^{1/2}$\AA. 
This indicates that the distorted CDW structure in \ce{CsV3Sb5} is relatively closer to the pristine structure than \ce{NbSe2}, which plays a crucial role in stabilizing the phonons. 


Our finding provides an alternative view to the prevailing understanding that the phonon softening behavior is a necessary feature for EPC-induced CDW transitions, holding potential implication for other unconventional CDW materials, such as cuprates, which exhibit approximately 15\% softening across the CDW transition~\cite{miao2018incommensurate,le2014inelastic}.
Using our notion described above, cuprates are likely to display an intermediate zero-point ensemble distribution between the ones of complete-softening \ce{NbSe2} and no-softening \ce{CsV3Sb5}. 

\subsection{Experimental Implications}


The lack of phonon softening in \ce{CsV3Sb5} also presents a significant deviation from the typical second-order phase transitions seen in conventional CDW materials. 
The phase transition in CDW materials is typically characterized by complete phonon softening, implying a continuous second-order transition.
The softening behavior indicates that the curvature of the free energy landscape profile at the pristine structure, which reflects the phonon stability, decreases with decreasing temperatures.
In the framework of Landau theory, this behavior indicates the continuous evolution of the pristine structure in the free energy landscape from a local minimum to a local maximum, implying a gradual modification of the order parameter from the pristine to the CDW phase.


In stark contrast, \ce{CsV3Sb5} exhibits a weak first-order phase transition, as evidenced by previous experiments, such as nuclear magnetic resonance~\cite{song2022orbital,luo2022possible}, elastic X-ray diffraction~\cite{li2022discovery}, heat capacity measurements~\cite{ortiz2020cs}, and thermal hysteresis behavior~\cite{jin2024pi}, 
etc. 
This behavior has previously been attributed to symmetry arguments, which allow for the presence of third-order terms in Landau theory~\cite{miao2021geometry}.
Our findings provide a physical explanation for the local stability of the pristine phase, which is stabilized by zero-point motion. 
As discussed in the previous section, we identified the dynamic stability of pristine \ce{CsV3Sb5} even below the CDW transition temperature. 
The stable pristine phase consistently represents an effective local minimum in the free energy landscape sketched in Fig.~\ref{fig:1-2}(a,b,c). 
The transition to the CDW phase is not accompanied by the gradual softening behavior. 
Instead, it occurs via a sudden shift in the order parameter indicated by the red arrow in Fig.~\ref{fig:1-2}(b), characteristic of a first-order phase transition.

Additionally, the influence of quantum fluctuations implies that the first-order phase transition in \ce{CsV3Sb5} is likely weak. 
The zero-point ensemble samples a broad range of distances, leading to an effective potential landscape that is considerably smeared.
Thus, the prominence of DFT energy barriers is reduced, further supporting the notion of a weak first-order transition.
An immediate implication of CDW being the first-order transition is that the pristine and CDW phases coexist as distinct local minima near the transition temperature, a phenomenon also detected in CPS with a long-lifetime metastable state~\cite{ratcliff2021coherent} and nuclear magnetic resonance with a narrow coexistence temperature range 91~K - 94~K~\cite{mu2021s}.
Another possible implication essentially suggests that the reduced energy barriers can be dynamically overcome, and thus we anticipatesmearing effects in the X-ray diffraction measurements in \ce{CsV3Sb5}~\cite{kautzsch2023incommensurate}.

\begin{figure}[!thb]
\includegraphics[width=1.0\textwidth]{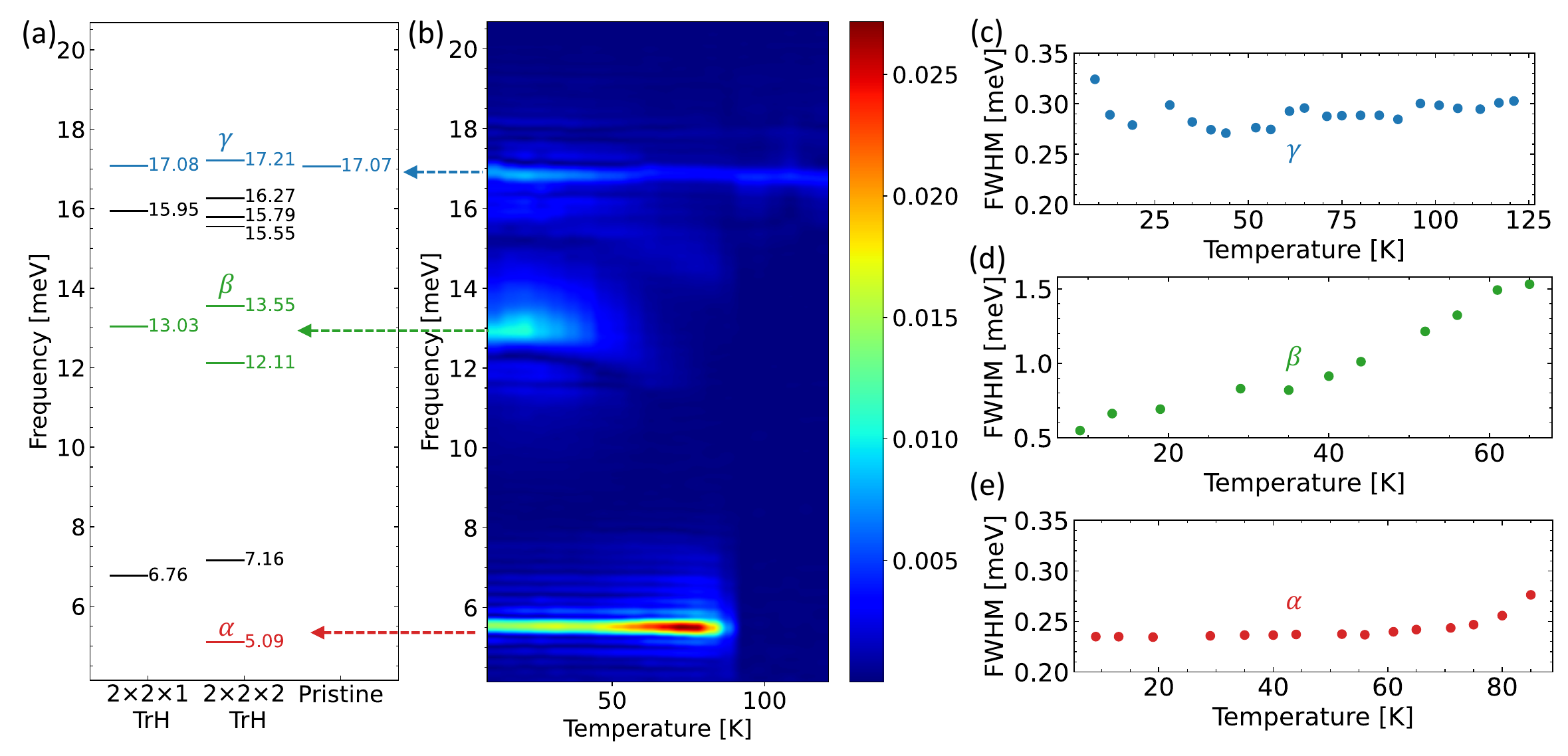}
\caption{
(a) Fully-symmetric phonon modes at $\Gamma$ point for the pristine 2$\times$2$\times$1 and 2$\times$2$\times$2 TrH orders.
(b) Coherent phonon spectroscopy for \ce{CsV3Sb5}.
The evolution of full-width half maximum (FWHM) with temperatures for the (c) $\gamma$, (d) $\beta$, and (e) $\alpha$ modes.
} 
\label{fig:1-4}
\end{figure}

We have also attempted to match the computational results with our CPS measurement of \ce{CsV3Sb5}.
Since CPS only observes the fully symmetric phonon mode, Fig.~\ref{fig:1-4}(a) includes the fully-symmetric phonon modes at the $\Gamma$ point of phonon dispersion for the \ce{CsV3Sb5} pristine, 2$\times$2$\times$1 TrH, and 2$\times$2$\times$2 TrH orders (see Supplemental Material~\cite{SM} for the irreducible representations of $\Gamma$ optical modes for 2$\times$2$\times$1 and 2$\times$2$\times$2 TrH orders).
The blue, green, and red colors are the $\gamma,\beta,\alpha$ modes aligned with the CPS result.
These modes have been previously reported in Ref.~\cite{ratcliff2021coherent} from the $M$ and $L$ point of the pristine phonon dispersion.

The $\gamma$ mode at 17.0~meV(4.1~THz), present at all temperatures, corresponds to the out-of-plane motion of the Sb atoms within the unit cell, oscillating toward and away from the Kagome plane~\cite{ratcliff2021coherent}. 
The $\alpha$ and $\beta$ modes are induced by symmetry breaking as the system enters the CDW phases. 
The $\alpha$ mode at 5.5~meV(1.3~THz) emerges at 94~K and corresponds to the 5.09~meV(1.23~THz) mode in the 2$\times$2$\times$2 TrH structure, supporting the transition picture described in Fig.\ref{fig:1-2}(b), where the 2$\times$2$\times$2 TrH structure is the most stable phase that first occurs at the transition temperature. 
The $\beta$ mode at 13.0~meV(3.1~THz) is broader and is observed only at lower temperatures around 60~K.
This mode likely arises from the competition between different CDW structures. 
Specifically, the 2$\times$2$\times$1 TrH structure has a mode at 13.03~meV(3.15~THz), while the 2$\times$2$\times$2 TrH structure has modes at 13.55~meV and 12.11~meV (3.28~THz and 2.93~THz).
This suggests that the metastable CDW structures, possibly including the 2$\times$2$\times$4 structures, have a set of modes with similar frequencies collectively leading to a broad $\beta$ spectrum.
The appearance of this $\beta$ mode at 60~K signals the onset of metastable states, reflecting the coexistence of multiple CDW minima, as depicted in Fig.~\ref{fig:1-2}(c).
Our analysis has also identified additional modes in black that are active in CPS due to CDW distortions. 
The correlation of these newly identified peaks with the experimental observations is notably strong.


The typical effect of decreasing temperature on the CPS active modes is a narrowing of the peak linewidths due to reduced electron-phonon and phonon-phonon scatterings.
However, as depicted in Figures~\ref{fig:1-4}(c,d,e), a saturation trend for the full-width at half maximum (FWHM) linewidth of the $\gamma,\beta,\alpha$ peaks emerges at lower temperatures (See Supplemental Material~\cite{SM} for peak positions, peak areas of the $\gamma,\beta,\alpha$ modes).
Our theory offers a potential explanation for this saturation behavior. 
Below the CDW transition temperature ($T<T_\mathrm{CDW}$), as illustrated in Fig.~\ref{fig:1-2}(c), the free energy profile introduces additional CDW orders as local minima with smeared energy barriers.
These local minima could act as a source of disorder through competing CDW phases, which would contribute to the broadening of phonon peak linewidths, counteracting the narrowing effect of a lowering temperature.
We propose that both the broadening and narrowing effects contribute to the observed saturation in the FWHM linewidth. 
This analysis elucidates the dynamical evolution of structural orders under thermal and zero-point effects, which is of fundamental importance in studying phase transitions in CDW materials.

It has been established that quantum fluctuations play a crucial role in driving quantum phase transitions in various materials, including cuprates~\cite{huang2021quantum} and \ce{NbSe2}~\cite{kundu2017quantum,soumyanarayanan2013quantum}, which occur near zero temperature where thermal effects are absent. 
It may seem paradoxical that quantum fluctuations can drive phase transition, but also impede the transition in  \ce{CsV3Sb5} by stabilizing the pristine phase.
We wish to highlight that both scenarios are possible depending on the specifics of the materials.
On one hand, quantum fluctuations may enable phase transitions by smearing energy barriers, making it easier for the system to undergo structural changes. 
On the other hand, when sufficiently strong, quantum fluctuations can completely smear out potential wells and tend to stabilize high-symmetry structures (Fig.~\ref{fig:1-2}d), thereby hindering transitions in some cases.
Other than \ce{CsV3Sb5}, it has been shown that in \ce{SrTiO3} the ferroelectric order is suppressed by quantum fluctuations~\cite{muller1979srti,verdi2023quantum}, and in 2H-\ce{NbSe2} quantum fluctuations can suppress the CDW phase transition at elevated pressures~\cite{leroux2015strong}.
Thus, the influence of quantum fluctuations on CDW transitions is complex, necessitating a detailed, case-by-case analysis.


\section{Conclusion}

In this study, we investigate the temperature-dependent energy landscape associated with CDW structures in \ce{CsV3Sb5} using \textit{ab initio} simulation, highlighting the important role of quantum zero-point motion.
We identify an energy hierarchy of CDW structures including two chiral 2$\times$2$\times$4 structures, with the 2$\times$2$\times$2 TrH order being the most energetically favorable.
Unlike typical CDW materials such as 2H-\ce{NbSe2}, the inclusion of zero-point motion results in the stabilization of phonons in pristine \ce{CsV3Sb5} even below the transition temperature.
In addition, quantum fluctuations can also smear the energy barriers, suggesting that \ce{CsV3Sb5} undergoes a weak first-order phase transition, accompanied by a coexistence of pristine and CDW phases near the transition temperature.
We propose a dynamic evolution of the free energy landscape, in which certain stable higher-energy CDW states, such as the 2$\times$2$\times$1 TrH order, could occur at lower temperatures.
Our theory aligns well with coherent phonon spectroscopy, particularly in the additional local minima states contributing to the saturating linewidth trend of the $\gamma$, $\beta$, and $\alpha$ modes.

Overall, this work challenges the conventional understanding that phonon softening must accompany EPC-induced CDW transitions.
\ce{CsV3Sb5} exemplifies how quantum mechanical effects can significantly influence macroscopic properties and the nature of phase transitions.
This research provides a novel perspective for further studies into other unconventional CDW materials and have rich implications in interpreting solid-state phase transitions.

\begin{acknowledgments}
The work is based on research supported via the University of California Santa Barbara (UCSB) Quantum Foundry funded via the National Science Foundation (NSF) Q-AMASE-i program under award DMR-1906325, and by the NSF Designing Materials to Revolutionize and Engineer our Future (DMREF) program under award DMR-2118523. 
This work used Stampede3 at Texas Advanced Computing Center (TACC) through allocation MAT200011 from the Advanced Cyberinfrastructure Coordination Ecosystem: Services \& Support (ACCESS) program, which is supported by National Science Foundation grants 2138259, 2138286, 2138307, 2137603, and 2138296. Use was also made of computational facilities purchased with funds from the National Science Foundation (award number CNS-1725797) and administered by the Center for Scientific Computing (CSC) at University of California, Santa Barbara (UCSB). The CSC is supported by the California NanoSystems Institute and the Materials Research Science and Engineering Center (MRSEC; NSF DMR-2308708) at UCSB. 
\end{acknowledgments}

\end{document}